\documentstyle[sprocl,epsf]{article}

\def\lsim{\mathrel{\raise.2ex\hbox{$<$}\hskip-.8em\lower.9ex\hbox{$\sim$}}}
\def\gsim{\mathrel{\raise.2ex\hbox{$>$}\hskip-.8em\lower.9ex\hbox{$\sim$}}}

\begin{document}


 \catcode`@=11
\def\@cite#1#2{\unskip\nobreak\relax
    \def\@tempa{$\m@th^{\hbox{\the\scriptfont0 #1}}$}%
    \futurelet\@tempc\@citexx}
\def\@citexx{\ifx.\@tempc\let\@tempd=\@citepunct\else
    \ifx,\@tempc\let\@tempd=\@citepunct\else
    \ifx;\@tempc\let\@tempd=\@citepunct\else
    \let\@tempd=\@tempa\fi\fi\fi\@tempd}
\def\@citepunct{\@tempc\edef\@sf{\spacefactor=\the\spacefactor\relax}\@tempa
    \@sf\@gobble}

\def\citenum#1{{\def\@cite##1##2{##1}\cite{#1}}}

\newcount\@tempcntc
\def\@citex[#1]#2{\if@filesw\immediate\write\@auxout{\string\citation{#2}}\fi
  \@tempcnta\z@\@tempcntb\m@ne\def\@citea{}\@cite{\@for\@citeb:=#2\do
    {\@ifundefined
       {b@\@citeb}{\@citeo\@tempcntb\m@ne\@citea\def\@citea{,}{\bf ?}\@warning
       {Citation `\@citeb' on page \thepage \space undefined}}%
    {\setbox\z@\hbox{\global\@tempcntc0\csname b@\@citeb\endcsname\relax}%
     \ifnum\@tempcntc=\z@ \@citeo\@tempcntb\m@ne
       \@citea\def\@citea{,}\hbox{\csname b@\@citeb\endcsname}%
     \else
      \advance\@tempcntb\@ne
      \ifnum\@tempcntb=\@tempcntc
      \else\advance\@tempcntb\m@ne\@citeo
      \@tempcnta\@tempcntc\@tempcntb\@tempcntc\fi\fi}}\@citeo}{#1}}
\def\@citeo{\ifnum\@tempcnta>\@tempcntb\else\@citea\def\@citea{,}%
  \ifnum\@tempcnta=\@tempcntb\the\@tempcnta\else
   {\advance\@tempcnta\@ne\ifnum\@tempcnta=\@tempcntb \else \def\@citea{--}\fi
    \advance\@tempcnta\m@ne\the\@tempcnta\@citea\the\@tempcntb}\fi\fi}
\catcode`@=12

\renewcommand{\theenumi}{\roman{enumi}}
\renewcommand{\labelenumi}{(\theenumi)}

\def\thefootnote{\fnsymbol{footnote}}

\font\fortssbx=cmssbx10 scaled \magstep1

\title{\vspace*{-.75in}
\hbox to \hsize{
\hbox{\fortssbx University of Wisconsin - Madison}
\hfill$\vtop{\normalsize\hbox{\bf MADPH-00-1211}
                \hbox{\rm December 2000}
                \hbox{\hfil}}$ }
\uppercase{Lepton Flavor Violating Era\\ of Neutrino Physics}\footnote{Talk presented at the {\it Joint U.S./Japan Workshop on New  
Initiatives in Lepton Flavor\break Violation and Neutrino Oscillation
with High Intense Muon and Neutrino Sources},\break Honolulu, Hawaii,
Oct.~2--6, 2000}}

\author{\vspace*{-.2in}\uppercase{V. Barger}}

\address{Physics Department, University of Wisconsin, Madison, WI 53706, USA}

\maketitle

\thispagestyle{empty}

\abstracts{
The physics agenda for future long-baseline neutrino oscillation experiments is outlined and the prospects for accomplishing those goals at future neutrino facilities are considered. Neutrino factories can deliver better reach in the mixing and mass-squared parameters but conventional super-beams with large water or liquid argon detectors can probe regions of the parameter space that could prove to be interesting.\vspace*{-3ex}}

\section{Introduction}

Neutrino oscillation phenomena probe the fundamental properties of neutrinos\cite{reviews}. We presently have evidence for (i)~atmospheric $\nu_\mu$ disappearance oscillations with mass-squared difference $\delta m^2_{\rm atm} \approx 3\times10^{-3}\rm\,eV^2$, (ii)~solar $\nu_e$ disappearance oscillations with $\delta m^2_{\rm solar} \approx 5\times10^{-5}\rm\,eV^2$, and (iii)~and accelerator $\bar\nu_\mu\leftrightarrow\bar\nu_e$ and $\nu_\mu\leftrightarrow\nu_e$ oscillations with $\delta m^2_{\rm LSND} \approx 1\rm~eV^2$. Limits from accelerator and reactor experiments place important constraints on oscillation possibilities. In particular, reactor experiments exclude large amplitude $\bar\nu_e$ disappearance oscillations at $\delta m^2 > 10^{-3}\rm\,eV^2$.

A 3-neutrino model can explain the atmospheric and solar data and provides a useful benchmark for neutrino factory studies. The mixing of 3 neutrinos can be parametrized by 3 angles ($\theta_{23},\ \theta_{12},\ \theta_{13}$) and a CP-violation phase ($\delta$). The angle $\theta_{23}$ controls the atmospheric oscillation amplitude, $\theta_{12}$ controls the solar oscillation amplitude, and $\theta_{13}$ couples atmospheric and solar oscillations and controls the amount of $\nu_e$ oscillations to $\nu_\mu$ and $\nu_\tau$ at the atmospheric scale.

What we now know from experiments is that:

\begin{enumerate} \addtolength{\itemsep}{-1ex}

\item $\theta_{23} \sim \pi/4\,,\ \left| \delta m_{32}^2 \right| \sim 3 \times 10^{-3}\rm\,eV^2$ for atmospheric oscillations;

\item $\theta_{12} \sim \pi/4\,, \ \left| \delta m_{32}^2 \right| \sim 5 \times 10^{-5}\rm\,eV^2$ is favored for solar oscillations (the LAM solution) but other values are not fully excluded;

\item $\theta_{13} \sim 0 \ (\sin^22\theta_{13} < 0.1)$ from the reactor experiments.

\end{enumerate}
In the limit $\theta_{13} = 0$, the oscillations are bimaximal\cite{bimax}.

A new round of accelerator experiments with medium baselines is under way\cite{expts}. The K2K experiment ($L=250$~km, $\left< E_\nu \right> \sim 1.4$~GeV) is finding evidence in line with the atmospheric $\nu_\mu$ disappearance. The MINOS experiment ($L=730$~km, $\left< E_\nu \right> \sim 10$~GeV) and the CNGS experiments ICANOE and OPERA ($L=730$~km, $\left< E_\nu \right> \sim 20$~GeV) are expected to ``see" the first oscillation minimum in $\nu_\mu\leftrightarrow\nu_\mu$, measure $\sin^22\theta_{23}$ to 5\% and $\delta m^2_{\rm atm}$ to 10\% accuracy, and search for $\nu_\mu\to\nu_e$ down to 1\% in amplitude. The short-baseline MiniBooNE experiment at Fermilab will confirm or reject the LSND effect\cite{lsnd}. However, information about neutrino masses and mixing will still be incomplete. Higher intensity beams of both $\nu_\mu$ and $\nu_e$ flavors are needed.

Conventional neutrino beams based on $\pi,\ K$ decays are dominantly $\nu_\mu$ and $\bar\nu_\mu$. Neutrino factories would provide high intensity $\nu_\mu$ and $\bar\nu_e$ (or $\bar\nu_\mu$ and $\nu_e$) beams from muon decays. The $\nu_e$ and $\bar\nu_e$ beams would access neutrino oscillation channels that are otherwise inaccessible and are essential for eventual reconstruction of the neutrino mixing matrix.

\section{Neutrino Factory}

Collimated high-intensity neutrino beams can be obtained from decays of muons stored in an oval ring with straight sections\cite{geer-ring}. For $\mu^-\to\nu_\mu e^-\bar\nu_e$ decays, the oscillation channels $\nu_\mu\to\nu_\mu$ (disappearance), $\nu_\mu\to\nu_e$ (appearance), and $\nu_\mu\to\nu_\tau$ (appearance) give ``right-sign" leptons $\mu^-,\ e^-$, and $\tau^-$, respectively, whereas the oscillation channels $\bar\nu_e\to\bar\nu_e$ (disappearance), $\bar\nu_e\to\bar\nu_\mu$ (appearance), and $\bar\nu_e\to\bar\nu_\tau$ (appearance) give ``wrong-sign" leptons $e^+,\ \mu^+$, and $\tau^+$, respectively. The oscillation signals are relatively background free. The charge-conjugate channels can be studied in $\mu^+$ decays. 

With stored muon energies $E_\mu \gg m_\mu$, the neutrino beam is highly collimated and its flux is $\Phi \simeq N(E_\mu/m_\mu)^2 / (\pi L^2)$, where $N$ is the number of useful muon decays and $L$ is the baseline. The $\nu N$ cross section rises linearly with $E_\nu$ and hence with $E_\mu$. The event rate is proportional to $(E_\mu)^3$. The $\bar\nu N$ cross section is about 1/2 of the $\nu N$ cross section. The charged-current cross section for $\nu_\tau N$ suffers from kinematic suppression at low neutrino energies. 

Muons are the easiest to detect. The sign of the muon needs to be measured to distinguish $\nu_\mu\to\nu_\mu$ (right-sign $\mu$) and $\bar\nu_e\to\bar\nu_\mu$ (wrong-sign $\mu$). The backgrounds to the wrong-sign signal are expected to be small if the energy of the detected $\mu$ is $\gsim4$~GeV, which requires stored muon energies $E_\mu \gsim 20$~GeV.
The sign of electrons is more difficult to determine. It might only be possible to measure the combined $\nu_\mu\to\nu_e$ and $\bar\nu_e\to\bar\nu_e$ events.
Tau-leptons can be detected kinematically or by kinks. The $\tau$-production threshold of $E_\nu = 3.5$~GeV requires higher energy neutrino beams. 

The critical parameters of neutrino factory experiments are the number of useful muon decays $N$, the baseline $L$, and the data sample size (kt-years), where the latter is defined as the product of the detector fiducial mass, the efficiency of the signal selection requirements, and the number of years of data taking. An entry-level machine may have $N = 6\times10^{19}$ and $E_\mu = 20$~GeV, while a high-performance machine may have $N = 6\times10^{20}$ and $E_\mu = 35$--70~GeV. The average neutrino beam energies are related to the stored muon energies by $\left< E_{\nu_\mu} \right> = 0.7 E_\mu$ and $\left< E_{\nu_e} \right> = 0.6E_\mu$. Baseline distances from 730~km to 10,000~km are under consideration. For an iron scintillator target a detector mass of 10--50~kt may be employed. With these factories, thousands of neutrino events per year could be realized in a 10~kt detector anywhere on Earth.
A number of recent studies have addressed the potential of long-baseline experiments to determine the neutrino masses and mixing parameters$^{6{-}14}$. 

\section{Physics Agenda (PA)}

There is a well-defined set of physics goals for long-baseline neutrino experiments, as follows.

\noindent{\bf PA1}: The measurement of $\theta_{13}$ is a primary goal. A nonzero value of $\theta_{13}$ is essential for CP violation and for matter effects with electron-neutrinos. The flavor-changing vacuum probabilities in the leading-oscillation approximation are
\begin{eqnarray}
P(\nu_e\to\nu_\mu) &\simeq& \sin^22\theta_{13} \sin^2\theta_{23} \sin^2 \left(\delta m_{\rm atm}^2 L\over 4E\right) \;,\\
P(\nu_e\to\nu_\tau) &\simeq& \sin^22\theta_{13} \cos^2\theta_{23} \sin^2 \left(\delta m_{\rm atm}^2 L\over 4E\right) \;,\\
P(\nu_\mu\to\nu_\tau) &\simeq& \cos^4 \theta_{13} \sin^22\theta_{23} \sin^2 \left(\delta m_{\rm atm}^2 L\over 4E\right) \;.
\end{eqnarray}
The $\nu_e\to\nu_\mu$ and $\nu_e\to\nu_\tau$ appearance channels provide good sensitivity to $\theta_{13}$; including the disappearance channels improves the sensitivity. All baselines are okay for a $\theta_{13}$ measurement.

\noindent{\bf PA2}: The sign of $\delta m^2_{32}$ determines the pattern of neutrino masses (i.e., whether the closely spaced mass-eigenstates that give $\delta m_{\rm solar}^2$ lie above or below the third mass-eigenstate). The coherent scattering of electron neutrinos in matter gives a probability difference $P(\nu_e\to\nu_\mu) - P(\bar\nu_e\to\bar\nu_\mu)$ that is positive for $\delta m^2_{32} >0$ and negative for $\delta m^2_{32} <0$. At baselines of about 2000~km or longer, a proof in principle has been given that the sign of $\delta m^2_{32}$ can be determined in this way at energies $E_\nu \sim 15$~GeV or higher\cite{bgrw}. A complicating factor is that fake CP violation from matter effects must be distinguished from intrinsic CP violation due to the phase $\delta$. 

In the presence of matter the $\nu_e\to\nu_\mu$ probability at small $\theta_{13}$ is approximately given by\cite{bgrw}
\begin{equation}
\left< P(\nu_e\to\nu_\mu) \right> \simeq {\sin^22\theta_{13}\over \left| 1 - {\left< A \right> \over \delta m^2_{32}} \right| } \sin^2 \left\{ 1.27 {\delta m_{32}^2 L\over \left< E_\nu \right>} \left| 1 - {\left< A \right> \over \delta m_{32}^2} \right| \right\} \,,
\end{equation}
where $\left< A \right> = 2\sqrt2 \, G_F \left< N_e \right> \left< E_\nu \right> $. The sign of $A$ is reversed for $\bar\nu_e\to\bar\nu_\mu$. Matter effects can enhance appearance rates by an order of magnitude at long baselines. One appearance channel is enhanced and the other suppressed so the separation of the $\nu_e\to\nu_\mu$ and $\bar\nu_e\to\bar\nu_\mu$ probabilities turns on as $L$ increases.

\noindent{\bf PA3}: Precision measurements of the leading-oscillation parameters at the few percent level are important for testing theoretical models of masses and mixing. The magnitude of $\delta m^2_{32}$ affects the shape of the oscillation suppression and $\sin^22\theta_{23}$ affects the amount of suppression, so both can be well measured by neutrino factories. 

\noindent{\bf PA4}: The subleading $\delta m^2_{\rm solar}$ oscillation can be probed if the currently favored large-angle mixing (LAM) solution to the solar neutrino problem proves correct. The KAMLAND reactor $\bar\nu_e$ experiment should also accurately measure the subleading-oscillation parameters of the LAM solution\cite{danny}.

\noindent{\bf PA5}: An important goal of neutrino factories is to detect intrinsic CP violation, $P(\nu_\mu\to\nu_e) \neq P(\bar\nu_\mu\to\bar\nu_e)$. This is only possible if the solar solution is LAM. Sensitivity to intrinsic CP violation is best for baselines $L=2000$--4000~km. Intrinsic CP violation at a neutrino factory dominates matter effects for small $\theta_{13}\ (\sim10^{-4})$, whereas matter effects dominate intrinsic CP for large $\theta_{13}\ (\sim10^{-1})$.

\section{Conventional Neutrino SuperBeams}

Conventional neutrino beams are produced from decays of charged pions. These beams of muon-neutrinos have small components of electron-neutrinos. Possible upgrades of existing proton drivers to megawatt (MW) scale are being considered to produce conventional neutrino superbeams\cite{upgrades}. An upgrade to 4~MW of the 0.77~MW beam at the 50~GeV proton synchrotron of the proposed Japan Hadron Facility (JHF) would give an intense neutrino superbeam (SuperJHF) of energy $E_\nu \sim 1$~GeV. An upgrade of the 0.4~MW proton driver at Fermilab would increase the intensity of the NuMI beam by a factor of four (SuperNuMI) with three options for the peak neutrino energy  [$E_\nu(\rm peak)\sim3$~GeV (LE), 7~GeV (ME), and 15~GeV (HE)]. The capabilities of these conventional superbeams to accomplish parts of the neutrino oscillation physics agenda are curently being explored\cite{upgrades,superbeam}. Very large water detectors or smaller liquid argon detectors with excellent background rejection would be used in conjunction with the superbeams.

\section{Physics Reach}

The results in this section summarize a recent comparative study\cite{superbeam} of superbeam and neutrino-factory physics capabilities in future medium- and long-baseline experiments.
The reach of various superbeam and neutrino factory options are compared in Table~\ref{tab:s}. In these results 3 years of neutrino running followed by 6 years of anti-neutrino running is assumed. For superbeams the argon detector (A) has 30~kt fiducial mass and the water detector (W) a 220~kt fiducial mass, a factor of 10 larger than SuperKamiokande; signal efficiency and estimated detector backgrounds are taken into account. For neutrino factories a 50~kt iron scintillator detector is assumed.

\begin{table}[h]
\def\arraystretch{1}
\vspace*{-4ex}

\caption[]{\label{tab:s}
Summary of the $\sin^2 2\theta_{13}$ reach (in units of $10^{-3}$) 
for various combinations of neutrino beam,
distance, and detector for 
(i) a $3\sigma$ $\nu_\mu \to \nu_e$ appearance with $\delta m^2_{21}
= 10^{-5}$~eV$^2$,
(ii) an unambiguous $3\sigma$ determination of the sign of $\delta m^2_{32}$ with $\delta m^2_{21} = 5\times10^{-5}$~eV$^2$ , and
(iii) a $3\sigma$ discovery of CP violation for
$\delta m^2_{21} = 5\times10^{-5}$, $1\times10^{-4}$, and $2\times10^{-4}\rm\ eV^2$, from left to right respectively. Dashes in the sign of $\delta m^2_{32}$ column
indicate that the sign is not always determinable. Dashes in the CPV
columns indicate CPV cannot be established for $\sin^22\theta_{13}
\le 0.1$, the current experimental upper limit, for any values of the
other parameters. The CPV entries are calculated assuming the value of
$\delta$ that gives the maximal disparity of $N(e^+)$ and $N(e^-)$; for
other values of $\delta$, CP violation may not be measurable.}
\medskip
\centering\leavevmode
\begin{tabular}{|ccc|c|c|ccc|}
\hline\hline
&&& \multicolumn{5}{c|}{$\sin^2 2\theta_{13}$ reach (in units of $10^{-3}$)}\\
\cline{4-8}
 Beam&  $L$ (km) &  Detector & (i) 
& (ii) & \multicolumn{3}{c|}{(iii)}\\
\hline\hline
JHF  & 295 & A & $25$ & $-$ & $-$ &  $-$ & $25$\\
     &     & W & $17$ & $-$ & $-$ & $40$ & $8$\\
\hline\hline
SJHF & 295 & A &  $8$ & $-$ &   $-$ &  $5$ & $3$\\
     &     & W & $15$ & $-$ & $100$ & $20$ & $5$\\
\hline\hline
SNuMI LE & 730 & A &  $7$ & $-$ & $100$ & $20$ &  $4$\\
         &     & W & $30$ & $-$ &   $-$ &  $-$ & $40$\\
\hline\hline
SNuMI ME & 2900 & A & $3$ & $6$ & $-$ & $-$ & $100$\\
         &      & W & $8$ & $15$ & $-$ & $-$ &   $-$\\
\hline
         & 7300 & A &  $6$ & $6$ & $-$ & $-$ & $-$\\
         &      & W &  $3$ & $3$ & $-$ & $-$ & $-$\\
\hline\hline
SNuMI HE & 2900 & A &  $3$ & $7$ & $-$ & $100$ & $20$\\
         &      & W & $10$ & $15$ & $-$ &   $-$ &   $-$\\
\hline
         & 7300 & A &  $4$ & $4$ & $-$ & $-$ & $-$\\
         &      & W &  $3$ & $3$ & $-$ & $-$ & $-$\\
\hline\hline
20 GeV NuF                 & 2900 & 50~kt & $0.5$ & $2.5$ & $-$ & $2$ & $1.5$\\
$1.8\times10^{20}$ $\mu^+$ & 7300 &       & $0.5$ & $0.3$ & $-$ & $-$ & $-$\\
\hline
20 GeV NuF             & 2900 & 50~kt & $0.1$ & $1.2$ & $0.6$ & $0.4$ & $0.6$\\
$1.8\times10^{21}$ $\mu^+$ & 7300 &   & $0.07$ & $0.1$ & $-$  & $-$   & $-$\\
\hline\hline
\end{tabular}
\end{table}

The $\sin^22\theta_{13}$ reach at neutrino factories depends on the subleading scale $\delta m_{21}^2$. This dependence is illustrated in Fig.~1 for stored muon energies of 20, 30, 40 and 50~GeV.

\begin{figure}[h]
\centering\leavevmode
\epsfxsize=3.15in\epsffile{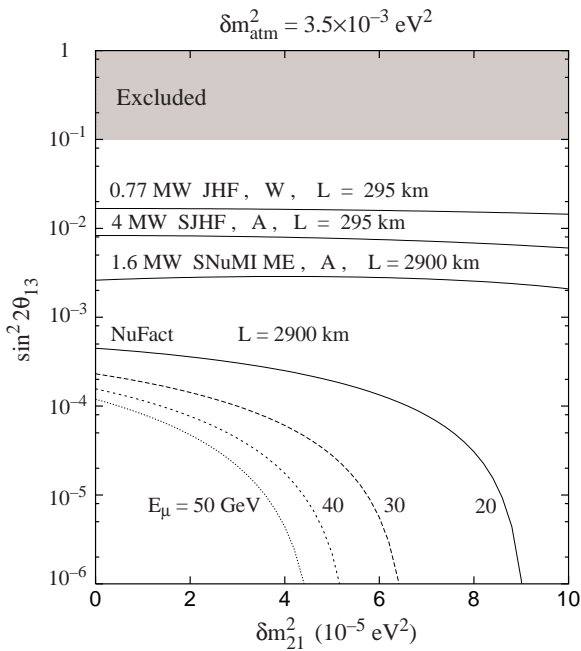}

\caption[]{Limiting $\sin^22\theta_{13}$ sensitivity for the observation of $\nu_\mu\to\nu_e$ oscillations expected with superbeams and neutrino factories versus the subleading scale $\delta m_{21}^2$. (Adapted from the study in Ref.~\citenum{superbeam}.)}
\end{figure}

The sign($\delta m_{32}^2$) and CP sensitivities are illustrated in Figures~2--5 for various neutrino beam and detector choices:
\begin{enumerate}
\addtolength{\itemsep}{-1mm}
\item neutrino factory with $E_\mu = 20$~GeV and $L = 2900$~km (Fig.~2);

\item SJHF with a water Cherenkov detector at $L=295$~km (Fig.~3);

\item SNuMI with an $E_\nu({\rm peak}) \sim 3$~GeV beam and a liquid argon detector at $L = 730$~km (Fig.~4);

\item SNuMI with an $E_\nu({\rm peak}) \sim 15$~GeV beam and a liquid argon detector at $L = 2900$~km (Fig.~5).

\end{enumerate}
In these SNuMI examples, the CP sensitivity is better in (iii) and the sign($\delta m_{32}^2$) sensitivity is better in (iv).

\begin{figure}[t]
\centering\leavevmode
\epsfxsize=2.9in\epsffile{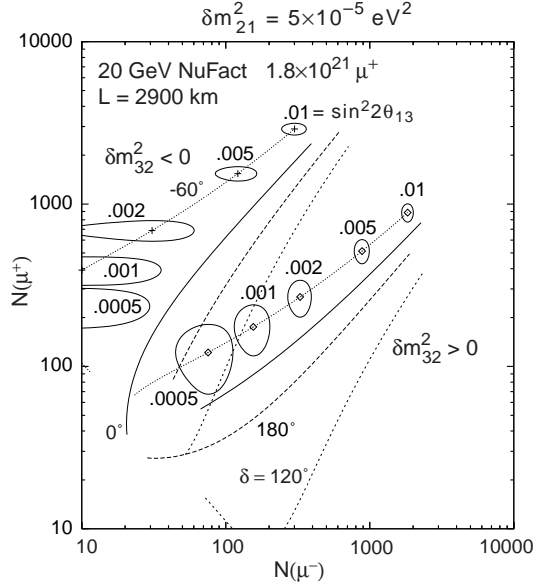}

\caption[]{$3\sigma$ error ellipses in the 
$\left[N(\mu^+), N(\mu^-)\right]$-plane, shown for a neutrino 
factory delivering $3.6\times10^{21}$ useful decays of 20~GeV muons and
$1.8\times10^{21}$ useful decays of 20~GeV antimuons, with a 50~kt detector
at $L = 2900$~km, for
$\delta m^2_{21} = 5\times10^{-5}$~eV$^2$. 
The solid and long-dashed curves correspond to the CP-conserving cases $\delta = 0^\circ$ and $180^\circ$, respectively, and the short-dashed and dotted curves correspond to two other cases that give the largest deviation from the
CP-conserving curves; along these curves $\sin^22\theta_{13}$ varies
from 0.0001 to 0.01, as indicated. (From Ref.~\citenum{superbeam}.)}
\end{figure}

\begin{figure}[t]
\centering\leavevmode
\epsfxsize=2.9in\epsffile{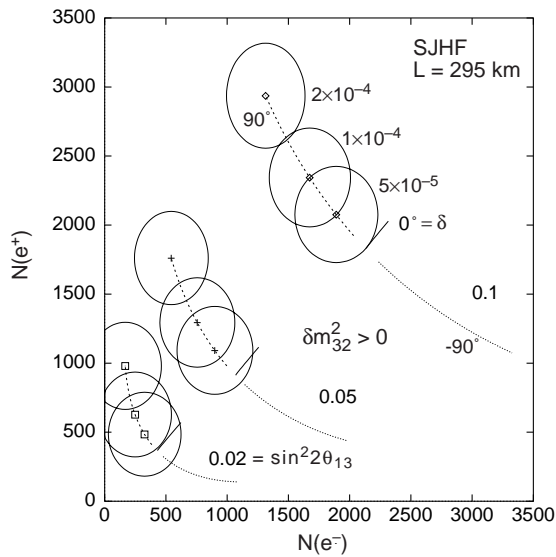}

\caption[]{$3\sigma$ error ellipses in the 
$\left[N(e^+), N(e^-)\right]$-plane, shown for the  4~MW SJHF scenario with 
$L = 295$~km. The contours are for the water Cherenkov detector scenario, with
$\sin^22\theta_{13} =
0.02$, $0.05$, and $0.1$. The solid (dashed) [dotted] curves correspond
to $\delta = 0^\circ$ ($90^\circ$) [$-90^\circ$] with $\delta m^2_{21}$
varying from $2\times10^{-5}$~eV$^2$ to $2\times10^{-4}$~eV$^2$.
The error ellipses are shown for three simulated data points
at $\delta m^2_{21} = 5\times10^{-5}$, $10^{-4}$ and
$2\times10^{-4}$~eV$^2$. (From Ref.~\citenum{superbeam}.)}
\end{figure}

\begin{figure}[t]
\centering\leavevmode
\epsfxsize=2.9in\epsffile{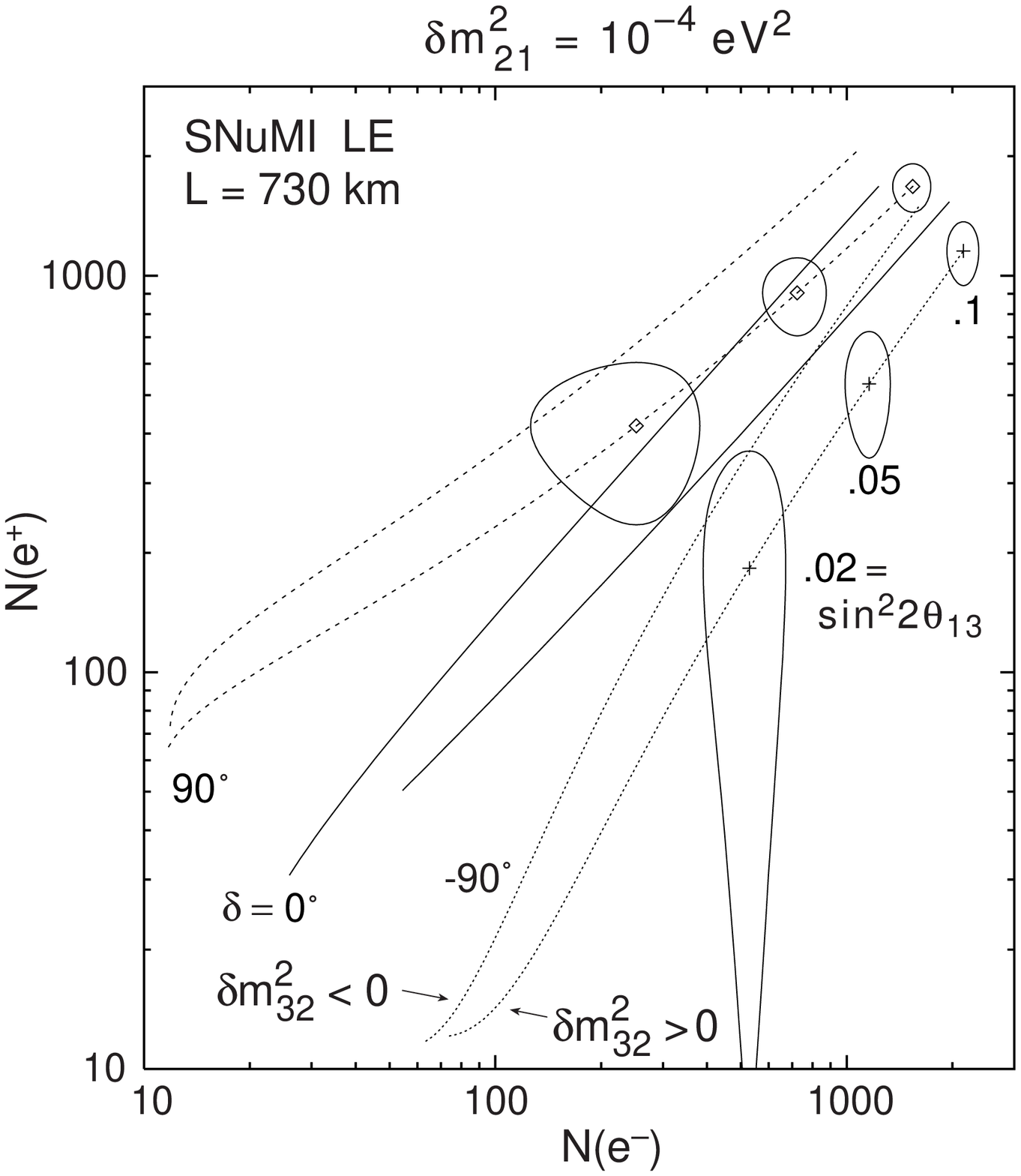}

\caption[]{$3\sigma$ error ellipses in the 
$\left[ N(e^+), N(e^-) \right]$-plane, shown for the liquid argon
detector scenario with the 
upgraded LE SNuMI beam at $L = 730$~km. 
The contours are for $\delta m^2_{21} = 10^{-4}$~eV$^2$. 
The solid and long-dashed curves 
correspond to the CP-conserving cases $\delta = 0^\circ$ and
$180^\circ$, respectively, and the short-dashed and dotted curves correspond to two other cases that give the largest deviation from the CP-conserving
curves; along these curves $\sin^22\theta_{13}$ varies from 0.001 to
0.1, as indicated. (From Ref.~\citenum{superbeam}.)}
\end{figure}

\begin{figure}[t]
\centering\leavevmode
\epsfxsize=2.9in\epsffile{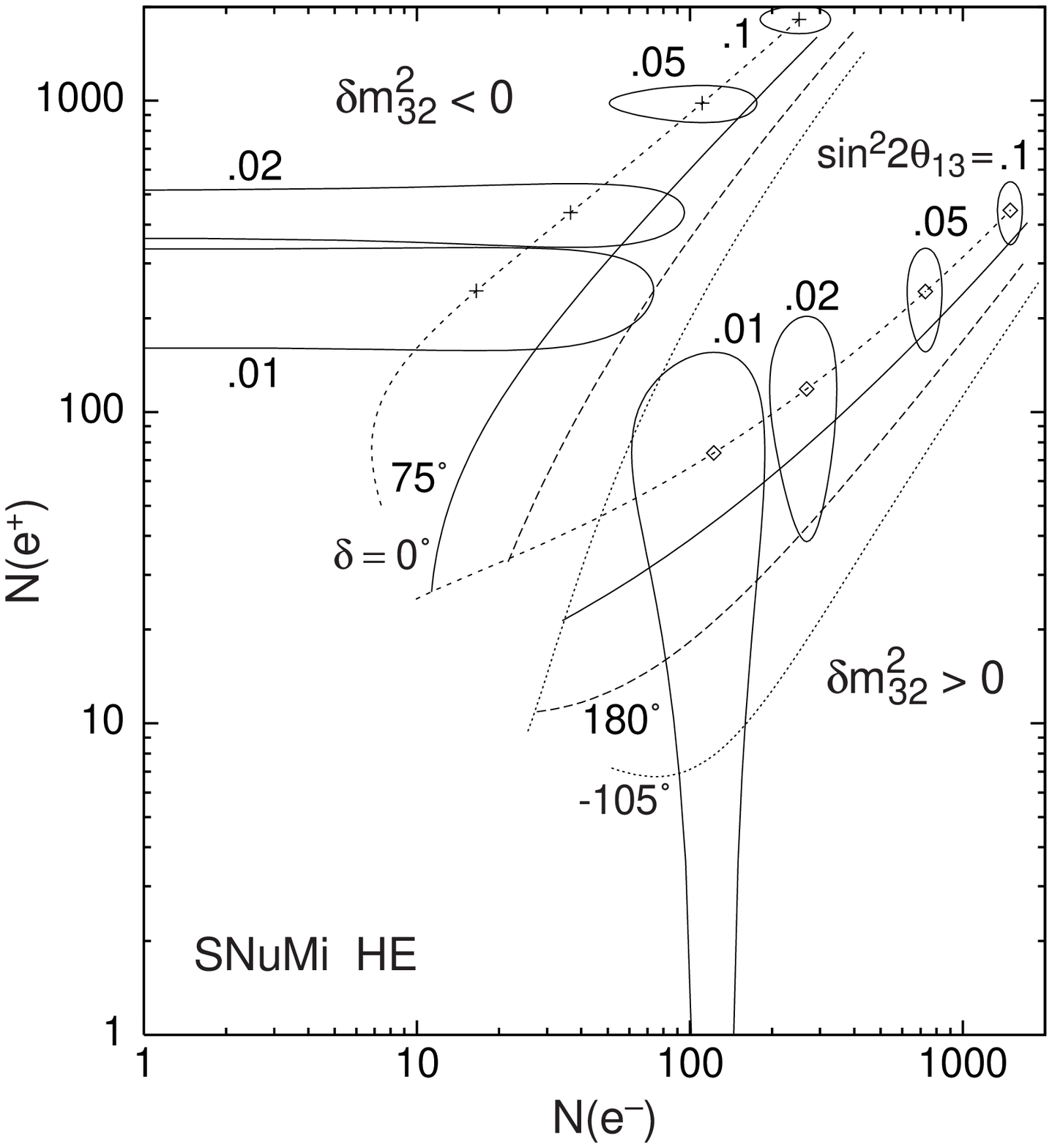}

\caption[]{$3\sigma$ error ellipses in the 
$\left[N(e^+), N(e^-)\right]$-plane, shown for the liquid argon
detector scenario at 
$L = 2900$~km with the 
upgraded HE SNuMI beam. 
The contours are for $\delta
m^2_{21} = 10^{-4}$~eV$^2$. The solid and long-dashed curves
correspond to the CP-conserving cases $\delta = 0^\circ$ and
$180^\circ$, respectively, and the short-dashed and dotted curves correspond to two other cases that give the largest deviation from the CP-conserving
curves; along these curves $\sin^22\theta_{13}$ varies from 0.001 to
0.1, as indicated. (From Ref.~\citenum{superbeam}.)}
\end{figure}

The conclusions from our comparative study of neutrino factories and conventional superbeams are as follows:

\begin{enumerate}
\addtolength{\itemsep}{-2mm}
\item A neutrino factory can deliver between one and two orders of magnitude better reach in $\sin^22\theta_{13}$ for $\nu_e\to\nu_\mu$ appearance, the sign of $\delta m^2_{32}$, and CP violation. At an $L=3000$~km baseline there is excellent sensitivity to all three observables. The $\sin^22
\theta_{13}$ reach is below $10^{-4}$. The sign of $\delta m^2_{32}$ can be determined and a detection of maximal CP violation made if $\sin^22\theta_{13}$ is larger that $10^{-3}$.

\item Superbeams with a sufficiently ambitious detector can probe $\sin^22\theta_{13}$ down to a few${}\times10^{-3}$. Maximal CP violation may be detected with a JHF or SJHF beam ($E_\nu\sim1$~GeV) at short baselines, but these facilities will have little sensitivity to sign($\delta m^2_{32}$). Higher-energy superbeams could determine sign($\delta m^2_{32}$) but have little sensitivity to CP violation.

\end{enumerate}

\section{Short Baselines}

If the LSND effect in $\nu_\mu\to\nu_e$ and $\bar\nu_\mu\to\bar\nu_e$ oscillations is confirmed by MiniBooNE, then an optimal baseline for future CP-violation studies with a neutrino factory would be
\begin{equation}
L \approx 45~{\rm km} \left( 0.3\ {\rm eV^2} \over \delta m^2_{\rm LSND} \right) \left( E_\mu\over 20\ \rm GeV\right) \,.
\end{equation}
The distance from Fermilab to Argonne is 30~km, for example. In four-neutrino oscillations, which would be indicated if the LSND, atmospheric, and solar effects are all due to neutrino oscillations, there are 3 CP-violating phases. The size of CP-violating effects in $\nu_e\to\nu_\mu$ and $\nu_\mu\to\nu_\tau$ may be enhanced or reduced relative to three-neutrino oscillations. 

\section{Overview}

We briefly sum up our conclusions regarding future neutrino factory and conventional superbeam studies of neutrino oscillations:

\begin{itemize}
\addtolength{\itemsep}{-2mm}
\item Three-neutrino mixing and $\delta m^2$ parameters can be measured at neutrino factories.

\item The amplitude $\sin^22\theta_{13}$ is the most crucial parameter. It can be measured down to $10^{-4}$ at a neutrino factory or to $3\times10^{-3}$ with superbeams.

\item A baseline $L \sim 3000$~km is ideal for neutrino factory measurements of sign($\delta m_{32}^2$), CP violation, and the subleading $\delta m_{21}^2$ oscillations.

\item A longer baseline, $L \sim 7300$~km, is best for precision on $\delta m_{32}^2$ and $\sin^22\theta_{32}$ at a neutrino factory; for superbeam measurements of $\sin^22\theta_{13}$ baselines of 3000--7000~km do equally well.

\item Measurements at two baselines  would provide complementary advantages (2800 and 7300 for neutrino factories or 295~km and 3000-7000~km for superbeams).

\item With the LAM solar solution, intrinsic CP-violating effects could be observable at SuperJHF and at a neutrino factory. The false CP-violation from matter is a serious but manageable complication at long baselines. Further studies are needed to determine the range of $\delta$ for which CP-violating effects are measurable.

\item For four-neutrino oscillations, short baselines ($L\sim 5$--50~km) are also important. With four neutrinos, CP violation occurs at the $\delta m_{\rm atm}^2$ scale and large effects may be seen.

\item Superbeams may be a reasonable next step in exploration of the neutrino sector. However, neutrino factories will eventually be needed for a complete understanding of the neutrino flavor-changing transitions.

\end{itemize}

\section*{Acknowledgments}

I thank K.~Whisnant for comments and S.~Geer, R.~Raja, and K.~Whisnant for collaboration on this study. This research was supported in part by the U.S.~Department of Energy under Grant No.~DE-FG02-95ER40896 and in part by the University of Wisconsin Research Committee with funds granted by the Wisconsin Alumni Research Foundation.

\end{document}